\documentclass[sigconf,nonacm]{acmart}

\usepackage[utf8]{inputenc}             
\usepackage[english]{babel}             

\newcommand{\etal}{et al.~}
\newcommand{\ie}{i.e.,~}
\newcommand{\eg}{e.g.,~}
\newcommand{\commentx}[1]{}

\setcopyright{rightsretained}

\begin{document}

\title{Gender Differences in Class Participation in Online versus In-Person Core CS Courses}

\author{Madison Brigham}
\email{mlbrigham@ucdavis.edu}
\affiliation{
  \department{Department of Computer Science}
  \institution{University of California, Davis}
  \city{Davis}
  \state{California}
  \country{USA}
  \postcode{95616}
}
\author{Joël Porquet-Lupine}
\orcid{0000-0003-4634-2877}
\email{jporquet@ucdavis.edu}
\affiliation{
  \department{Department of Computer Science}
  \institution{University of California, Davis}
  \city{Davis}
  \state{California}
  \country{USA}
  \postcode{95616}
}



\begin{abstract}

The COVID-19 pandemic significantly altered how post-secondary students receive their education.
Namely, the transition from an in-person to an online class format changed how students interact with their instructors and their classmates.
In this paper, we use student participation scores from two core computer science classes across ten in-person and three online quarters at a public research university to analyze whether the shift to primarily asynchronous online learning has impacted the gender gap in student participation scores and students' attitudes towards themselves and their peers.
We observe a shift on the online class forum: in in-person classes, males score higher on average and dominate the top scores while in online classes, male and female students participate at approximately the same rate classwide.
To understand what might be driving changes in participation behavior, we analyze survey responses from over a quarter of the students enrolled in the online classes.
While we find that students of both genders tend to compare themselves to their peers less when classes are online, we also find that this trend is much more accentuated for females than males.
This data suggests that observed female participation habits in typical in-person classes are not inherent gender differences, but rather, a product of the environment.
Therefore, it is critical the community investigates the root causes of these behavioral differences, and experiments with ways to mitigate them, before we soon return to an in-person format.

\end{abstract}

\maketitle

\section*{Forewords}

Despite our best efforts, this manuscript was rejected twice from top CS Education conferences (ACM SIGCSE TS and ACM ITiCSE) in 2022.
This version incorporates modifications from the latest round of feedback we received (limitations section, mention of non-binary responses, etc.).
Unfortunately, it was never resubmitted as Maddii graduated at the end of 2022, and the rationale for studying the effects of the pandemic was beginning to diminish.
We are making the decision to submit to arXiv as we strongly believe that this work is valuable and should reach its intended audience, even if not via peer-reviewed venues.


\section{Introduction}

The onset of the COVID-19 pandemic in early 2020 drastically affected the delivery of post-secondary education.
Across the US, colleges and universities rapidly moved their in-person instruction to an online format, which remained the norm at many institutions throughout the 2020 to 2021 academic year.
This widespread shift to online learning subsequently led to a decline in in-person lectures, where classroom climate is typically established.

As explained by Barker and Garvin-Doxas \cite{4,5}, classroom climate describes the communication and interaction patterns that shape the perceptions of students in a class.
A defensive classroom climate is characterized by judgement, apathy, and superiority.
This tends to result in students isolating from one another, feeling the need to compete with each other, and seeing experience-level as what constitutes intelligence.
This kind of culture can have an especially harmful impact on students in underrepresented groups, such as female students, whose enrollment in computer science (CS) programs has steadily declined since the mid-1980s \cite{2}.

In this paper, we investigate whether male and female students' attitudes and participation habits have changed with classes transitioning from in-person to online.
Does the shift from an in-person to an online class format impact the gender gap in student participation scores?
And has this shift also impacted students' attitudes towards themselves and their peers?

To measure this, we compared 1,860 student participation scores from ten in-person offerings of two core CS classes with 547 student participation scores from three --primarily asynchronous-- online offerings of the same classes, as explained in section \ref{sec:methodology}.
In section \ref{sec:scores}, we compare the scores received by male and female students in the two main participation categories, which remained consistent across class formats: participation on the class forum and survey completion.
In section \ref{sec:responses}, we evaluate students' perceptions of how their participation and view of themselves and their peers has or has not changed following the transition online.

\section{Related work}

The effects of the shift from in-person to online classes at the onset of the COVID-19 pandemic remain relatively unstudied, as the pandemic only began a little over a year and a half ago.
With that being said, a limited body of research has already been conducted in this area.
Work by Mooney and Becker \cite{6} found that male and female students who did not consider themselves to be part of a CS minority experienced a decline in their sense of belonging over the course of the pandemic.
Conversely, female students who considered themselves to be part of a minority saw an increase in their sense of belonging during this time period.
Work by Lewis \etal \cite{7} found that students reported their stress levels and perception of class difficulty in online classes to be similar to or lower than that of in-person classes, likely due in part to the increased accessibility of materials, the redesign of courses, and the increased flexibility in grading policies.
Our research aims to further the understanding of the effects of the shift to online schooling.
As the transition to online schooling subsequently changed how students interact with each other, their instructors, and the class material, we want to study potential changes in class participation and classroom climate, both well-established areas of research in terms of pre-pandemic schooling, under the lens of this shift.

Class participation in in-person classes can provide insight into the experience of students in CS.
Of particular interest is the experience of female students, who are underrepresented in CS at the post-secondary level \cite{2}.
Work by Alvarado \etal \cite{1} found that female students were less comfortable asking questions in lecture than males.
Similarly, work by Brigham and Porquet-Lupine \cite{3} found that the top class participation scorers in lecture and on class forums were disproportionately male.
Previous work has found that female students are more likely than males to ask and answer questions anonymously on the class forum \cite{8,9}.

Past research has found low self-confidence among female CS students \cite{20}.
Research by Winter \etal found that female students commonly felt as though they were falling behind their peers, and expressed concerns about their experience gaps with prior coding \cite{10}.
Even when quantitative ability is controlled for, female students express lower self-confidence with computing than their male counterparts \cite{1,13}.
Work by Treu and Skinner suggests that low self-confidence in female students can be attributed to negative stereotypes, subtle bias from instructors, and a lack of female role models \cite{14}.
Work by Cohoon also found that lower presence of female faculty was correlated with higher female attrition rates \cite{17}.

To combat this, efforts have been made to improve the self-perception of females in CS.
Research by Fisk \etal found that incorporating encouraging emails from the instructor to increase the self-perception of top-performing female students seemed to increase female intent to continue in CS \cite{11}.
Research by Krause-Levy \etal found that the design of open online computing classes that normalized struggle was correlated with higher female enrollment and higher completion rates for both male and female students \cite{12}.
Efforts have also been made to expose students to more female role models in the field \cite{15,19}, mitigate harmful side effects from experience gaps between students starting out in CS \cite{15,16,19}, and combat the stereotype that CS is isolated from other disciplines \cite{15,18}.

\section{Methodology}
\label{sec:methodology}

This study\footnote{Both the class data analysis and student survey in this study were reviewed by our university's Institutional Review Board.} looks at in-person and online offerings of two core classes that all CS majors at the university are required to take: Data Structures and Algorithms (hereafter referred to as \textbf{CS3}) and Operating Systems (hereafter referred to as \textbf{CSOS}).
Both of these quarter-long classes are concept-focused and programming-heavy.

CS3 is the third and final class in the introductory CS series, and students typically take it towards the end of their freshman year or beginning of their sophomore year.
For the in-person offerings of this course, we analyzed 333 students' grades across two quarters.
Of these students, 76.88\% were male and 23.12\% were female.
For the online offerings of CS3, we analyzed 153 students' grades across one quarter.
Of these students, 76.47\% were male and 23.53\% were female.
CSOS is an upper division course that students typically take towards the end of their junior year or beginning of their senior year.
For the in-person offerings of this course, we analyzed 1,504 students' grades across eight quarters.
Of these students, 75.40\% were male and 24.60\% were female.
For the online offerings of CSOS, we analyzed 392 students' grades across two quarters.
Of these students, 74.74\% were male and 25.26\% were female.

All classes were taught by the same male instructor, who collected the in-person class data from January 2017 to March 2020, and the online class data from April 2020 to March 2021.
Based on our university's registrar data, which uses a binary gender classification, students were classified as either male or female\footnote{We acknowledge that this is an oversimplified gender spectrum.}.
The 0.01\% of students (25 of the 2,407 students) whose gender was unknown by the registrar were omitted from our data analysis.

A student's participation score in CS3 and CSOS could typically earn them up to 3\% extra credit.
In the in-person offerings of these classes, this score was determined by three categories: lecture, forum, and survey.
For the online offerings of these classes, where lectures occurred asynchronously, the categories that determined a student's participation score slightly shifted: instead of earning points for participating in lecture, a student could earn points by attending synchronous office hours with either the instructor or TAs.
In this study, we focus on the two categories that remain consistent across class formats: forum and survey.

\paragraph{Forum participation} In both the in-person and online classes, a student earned points for forum participation by posting on Piazza, an online class forum where students can interact with one another in real time.
Students can post questions, answer other students' questions in a designated answer box on each post, or leave follow-up questions or comments in the discussion section at the bottom of each post.
Each student's forum score was determined by these three metrics, which are tracked by Piazza, with answers earning the most points, then questions, then follow-ups.
Students are able to post anonymously to classmates, but not to the instructor.

\paragraph{Survey participation} In both the in-person and online classes, a student earned points for survey participation by completing class evaluations throughout the quarter.
These include the official university course evaluation, as well as a mid-quarter and end-of-the-quarter survey created by the instructor.
Surveys provide opportunities for students to evaluate and leave feedback on the course, and students' responses to these surveys are seen only by the instructor.

\paragraph{Student input} To help us explain the trends we observed in the online offerings of CS3 and CSOS, we developed a short survey to collect student input.
This student survey was sent in February 2021 to the 533 students who had taken or were currently enrolled in CS3 online or CSOS online, so that we could capture responses specifically from the students composing our online class data set.
Participation in the survey was optional, and student responses were completely anonymous.
We received 138 complete responses, giving us a response rate of 25.89\%.
Of the students who responded, 71.01\% identified as male, 27.54\% identified as female, and 1.45\% identified as non-binary.
This leaves the gender ratio of the survey respondents close to that of the students composing the online class data set, suggesting that our survey responses are representative of our target population as a whole.
In this study, we primarily focus on male and female responses only because a sample size of two non-binary CS students is not sufficient to draw general conclusions about this gender category.

\section{Participation scores}
\label{sec:scores}

We began by comparing scores received in the two participation categories that remained consistent across class formats: the forum and survey participation.
For all statistical tests, we used an alpha level of .05.

\subsection{Forum}

\begin{figure}[h]
  \centering
  \includegraphics[width=0.45\textwidth]{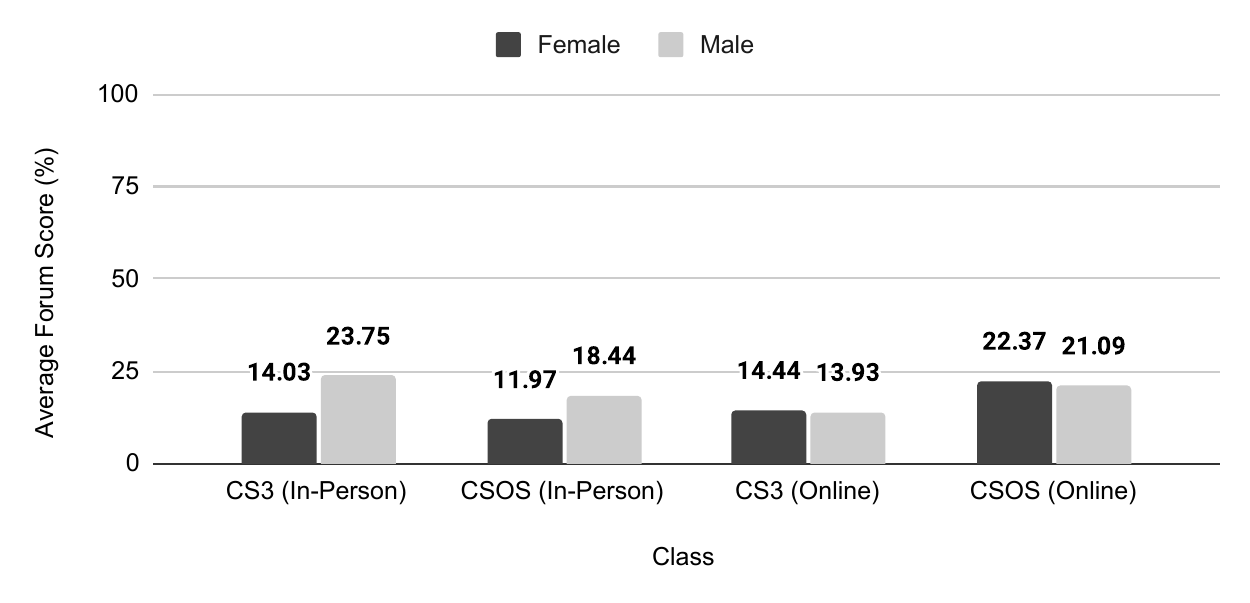}
  \caption{Average forum participation scores}
  \label{fig:forum_avgs}
\end{figure}

For CS3 in-person, the average male forum participation score (\textit{M} = 23.75, \textit{SD} = 32.79) was 1.69 times the average female forum participation score (\textit{M} = 14.03, \textit{SD} = 20.73), \textit{t}(331) = 3.11, \textit{p} = .002.
For CSOS in-person, the average male forum participation score (\textit{M} = 18.44, \textit{SD} = 26.70) was 1.54 times the average female forum participation score (\textit{M} = 11.97, \textit{SD} = 18.70), \textit{t}(1502) = 5.16, \textit{p} < .001.
Figure \ref{fig:forum_avgs} shows a visual comparison of these averages on the left.

In the online offerings of these courses, the forum participation imbalance disappeared.
For CS3 online, there is no longer a significant difference between the average male forum participation score (\textit{M} = 13.93, \textit{SD} = 21.99) and average female forum participation score (\textit{M} = 14.44, \textit{SD} = 23.45), \textit{t}(151) = -0.12, \textit{p} = .908.
For CSOS online, there is also no significant difference between the average male forum participation score (\textit{M} = 21.09, \textit{SD} = 28.92) and average female forum participation score (\textit{M} = 22.37, \textit{SD} = 27.29), \textit{t}(390) = -0.40, \textit{p} = .691.
Figure \ref{fig:forum_avgs} shows a visual comparison of these averages on the right.

While it is evident that the forum averages are much closer in the online classes than the in-person ones, we were curious about what was driving this change.
In other words, were a couple of female outliers pulling up the average for everyone, or had there been shifts in behavior classwide?
In order to measure this, we considered the distribution of participation scorers.

\begin{figure}[h]
  \centering
  \includegraphics[width=0.45\textwidth]{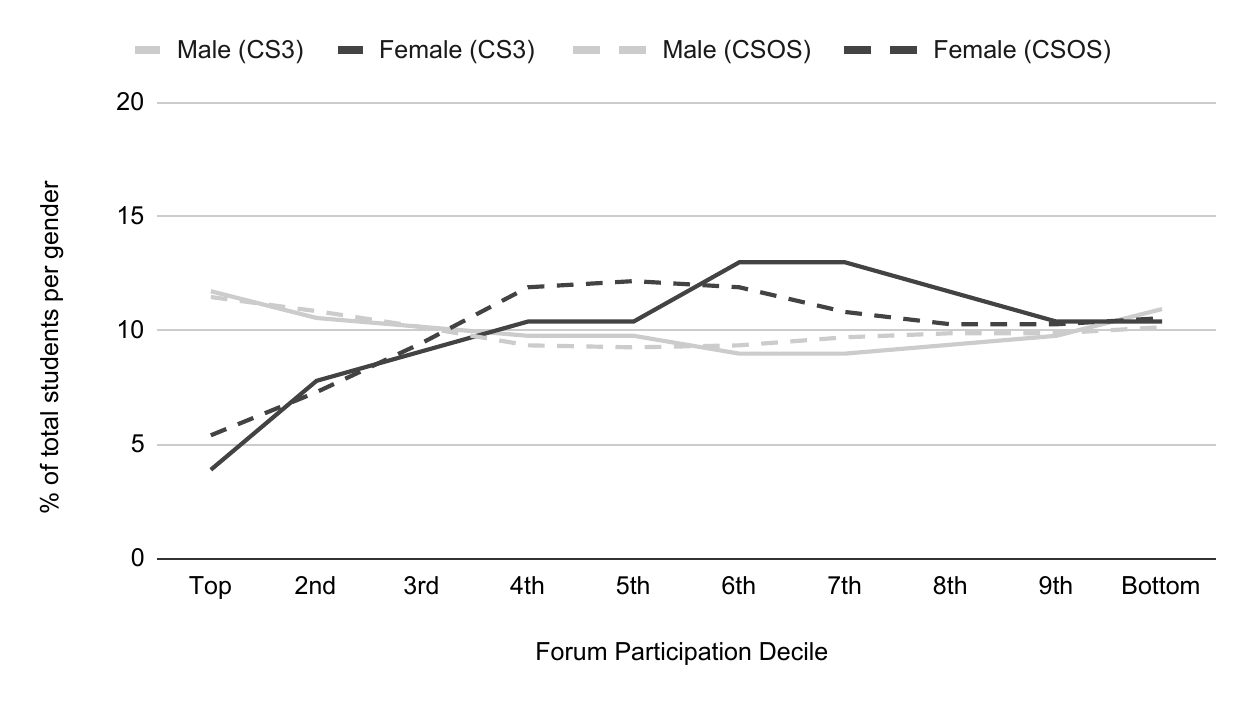}
  \caption{In-person distribution of forum scorers}
  \label{fig:forum_dist_inperson}
\end{figure}

\begin{figure}[h]
  \centering
  \includegraphics[width=0.45\textwidth]{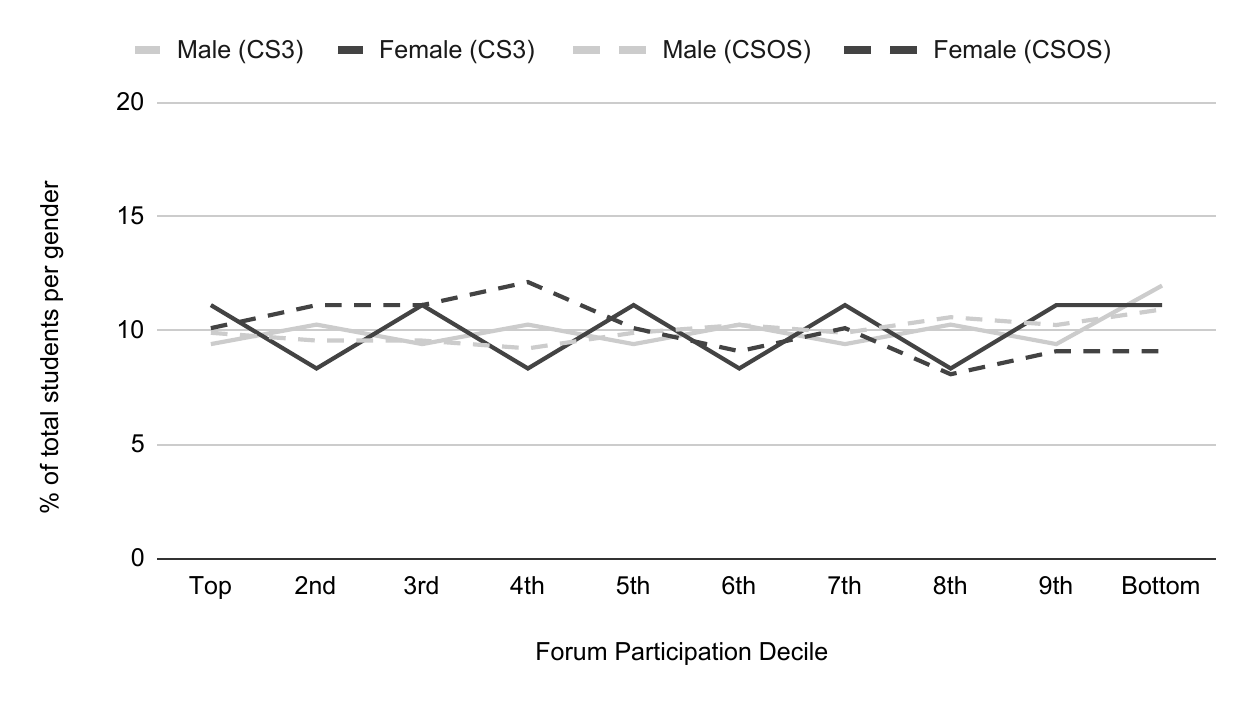}
  \caption{Online distribution of forum scorers}
  \label{fig:forum_dist_online}
\end{figure}

Decile distributions of forum participation scorers, like Figures \ref{fig:forum_dist_inperson} and \ref{fig:forum_dist_online}, help us visualize how male and female students rank in comparison to one another classwide, and therefore provide insight into this shift in forum averages.
When we combine the participation scores from all in-person or online quarters of a class, and rank them from highest to lowest, the distribution shows what percentage of male and female students scored in the top 10\% of scorers, the second decile of scorers, and so on down to the bottom 10\% of scorers.
If male and female students' scores were distributed evenly, each decile of scorers would contain 10\% of the female students and 10\% of the male students.

In the in-person offerings of CS3 and CSOS, which are shown in Figure \ref{fig:forum_dist_inperson}, we found a disproportionately low percentage of female students scoring in the top deciles of scorers, and a disproportionately high percentage of them scoring in the mid to low deciles.

In the online classes, which are shown in Figure \ref{fig:forum_dist_online}, we see a different pattern.
For both CS3 and CSOS online, male and female students participated on the forum at approximately the same rate classwide.
In other words, the proportion of male to female students who scored in each decile is approximately equal to the proportion of males to females enrolled in each class.
The dented nature of the CS3 female line can be attributed to small sample size.
The 35 females enrolled in CS3 online were as evenly distributed as possible, however, a difference of one student per decile (\eg 3 instead of 4) creates a misleadingly pronounced bend to the line.

Moving online seems to have closed the gender gap in average forum scores.
Furthermore, the average female forum score is not being pulled up by a handful of female students; instead, forum participation scores of male and female students are more evenly distributed classwide.

\subsection{Survey}

\begin{figure}[h]
  \centering
  \includegraphics[width=0.45\textwidth]{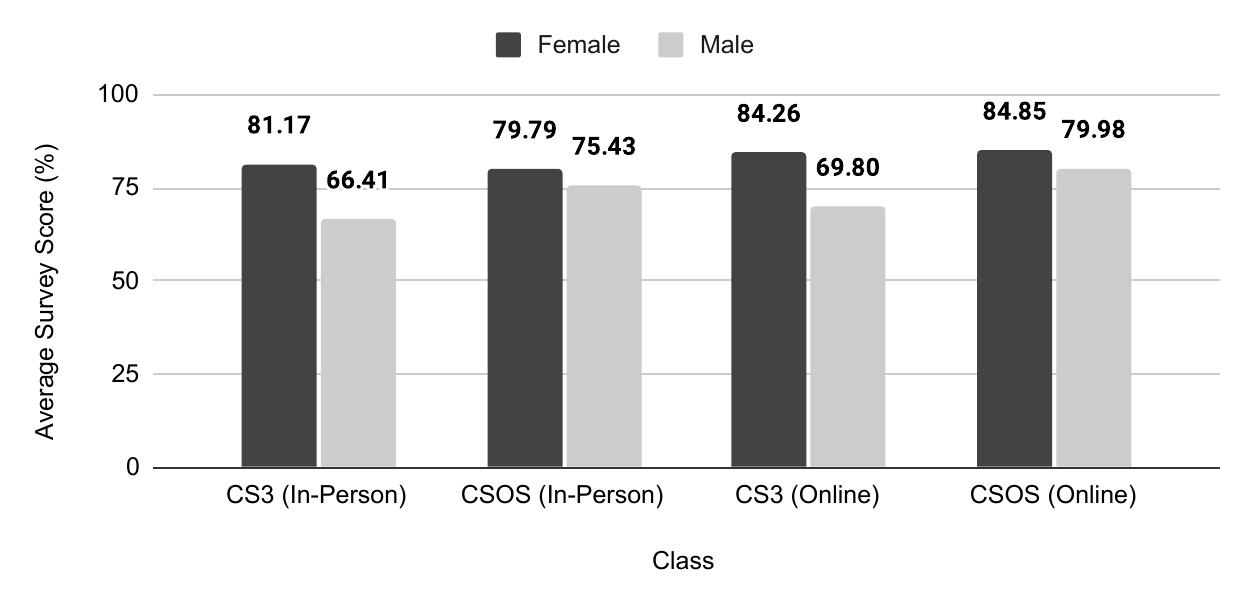}
  \caption{Average survey participation scores}
  \label{fig:survey_avgs}
\end{figure}

For CS3 in-person, the average female survey participation score (\textit{M} = 81.17, \textit{SD} = 32.48) was 1.22 times the average male survey participation score (\textit{M} = 66.41, \textit{SD} = 37.89), \textit{t}(331) = -3.36, \textit{p} = .001.
For CSOS in-person, the average female survey participation score (\textit{M} = 79.79, \textit{SD} = 31.47) was 1.06 times the average male survey participation score (\textit{M} = 75.43, \textit{SD} = 34.69), \textit{t}(1502) = -2.25, \textit{p} = .025.
Figure \ref{fig:survey_avgs} shows a visual comparison of these averages on the left.

In the online offerings of these courses, survey participation has mostly stayed the same.
For CS3 online, the average female survey participation score (\textit{M} = 84.26, \textit{SD} = 28.16) is 1.21 times the average male survey participation score (\textit{M} = 69.80, \textit{SD} = 38.90), \textit{t}(151) = -2.45, \textit{p} = .017.
For CSOS online, while there is no longer a statistically significant difference between the average male survey participation score (\textit{M} = 79.98, \textit{SD} = 30.86) and average female survey participation score (\textit{M} = 84.85, \textit{SD} = 25.32), \textit{t}(390) = -1.56, \textit{p} = .120, this may have to do with the smaller sample size of CSOS online compared to CSOS in-person.
Figure \ref{fig:survey_avgs} shows a visual comparison of these averages on the right.

\begin{figure}[h]
  \centering
  \includegraphics[width=0.45\textwidth]{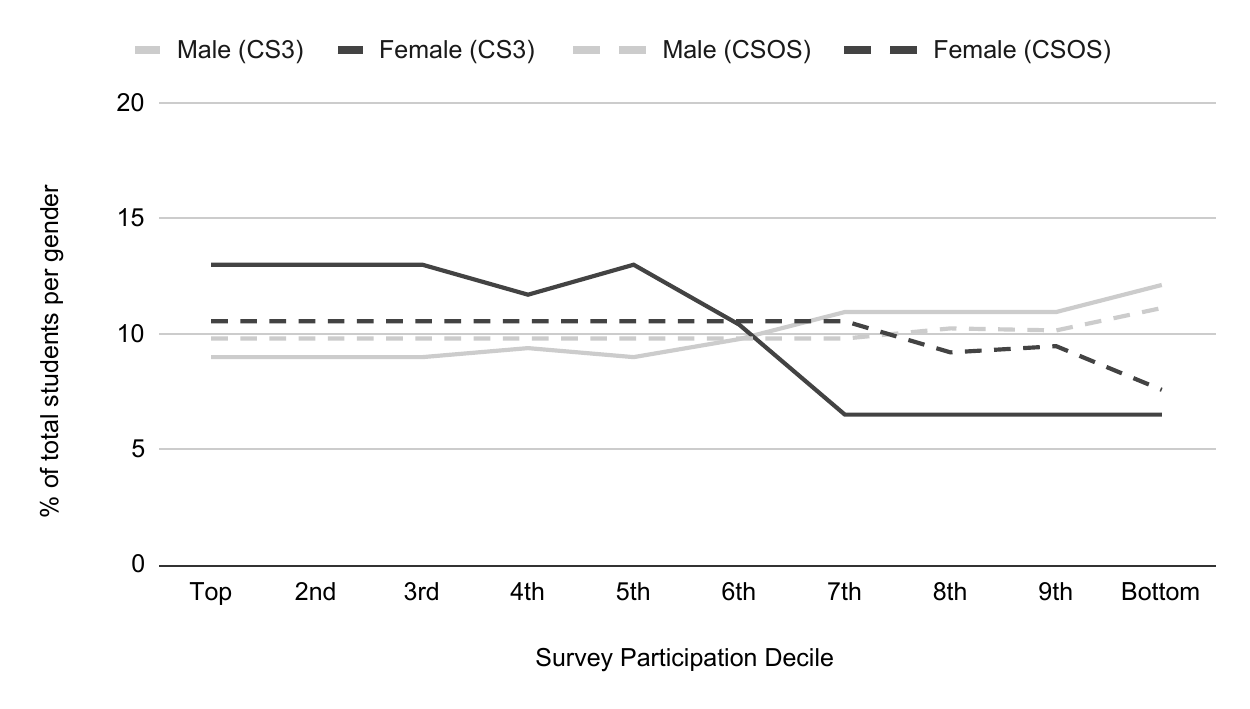}
  \caption{In-person distribution of survey scorers}
  \label{fig:survey_dist_inperson}
\end{figure}

\begin{figure}[h]
  \centering
  \includegraphics[width=0.45\textwidth]{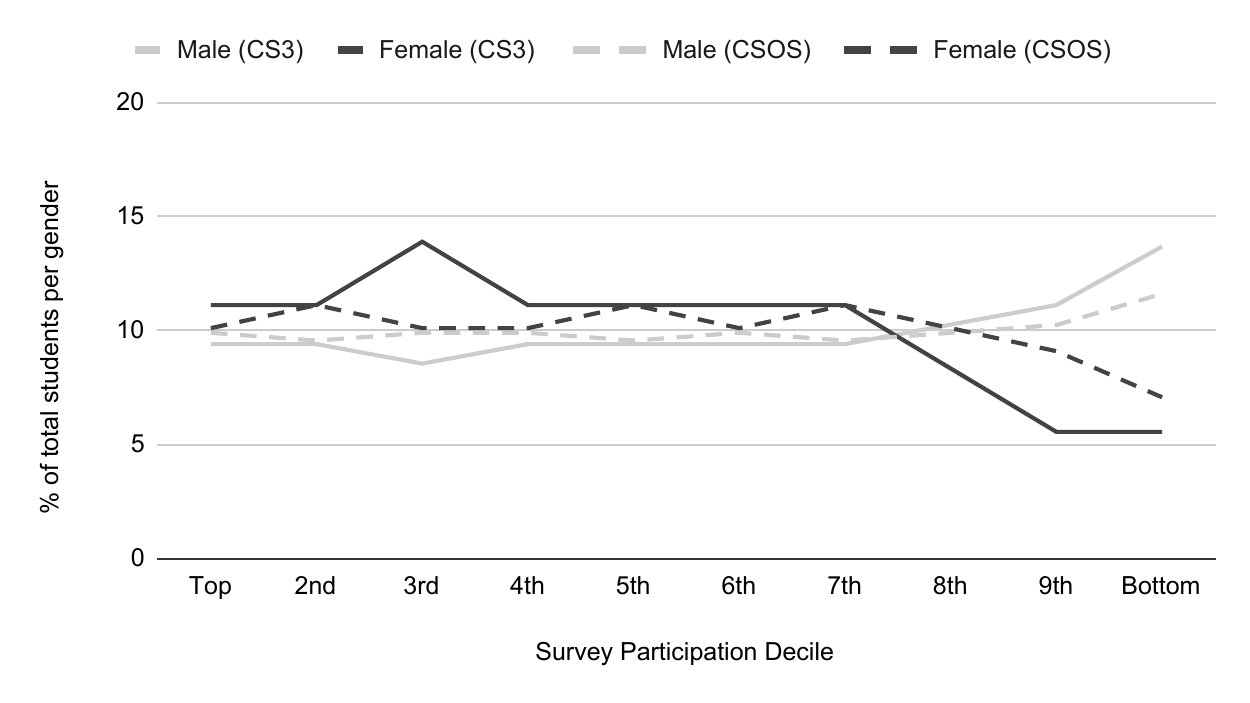}
  \caption{Online distribution of survey scorers}
  \label{fig:survey_dist_online}
\end{figure}

The decile distributions of survey participation scorers, shown in Figures \ref{fig:survey_dist_inperson} and \ref{fig:survey_dist_online}, reveal a similar pattern across class formats.
In both the in-person and online offerings of CS3 and CSOS, we consistently see a slightly higher percentage of female students scoring in the top and middle deciles, and a noticeably lower percentage of them scoring in the bottom deciles.

Based on these averages and distributions, survey participation patterns of male and female students appear to be relatively unchanged, despite the shift from an in-person to an online class format: the data suggests that in both in-person and online offerings of CS3 and CSOS, female students were more likely to complete the class surveys than their male counterparts.

\section{Student survey responses}
\label{sec:responses}

While analyzing class data can help us identify trends, collecting student input is vital in explaining them.
We wanted to know why globally, females are participating on the forum at the same rate as males now that classes are online, and whether this change in participation habits can be explained in terms of student attitudes.
To help us accomplish this, we developed a short student survey.

\subsection{Survey details}

In our survey, we wanted to assess two things: whether students believed that their forum participation had changed since going online (Question 1), and whether they believed that their attitudes had shifted as well.
We decided to capture the student attitudes which may have impacted participation on the forum by asking whether the shift from an in-person to an online class format had impacted students' confidence in their ability to succeed in their CS classes (Question 2), or the amount that they compared themselves to their peers (Question 3).

\textbf{Question 1:} Which statement best describes your forum participation in in-person versus online CS classes?
\textit{When classes are online, I participate \_\_\_ on the forum.}

\textbf{Question 2:} Which statement best describes your confidence in your own ability to succeed in your CS classes?
\textit{When classes are online, I feel \_\_\_ confident in my CS abilities.}

\textbf{Question 3:} Which statement best describes your attitudes towards your classmates?
\textit{When classes are online, I compare myself to my classmates \_\_\_.}

Students selected their response to each question from a five-point Likert scale: significantly more, slightly more, the same, slightly less, or significantly less.
Students could also respond to an optional free response following each question, if the student wished to explain their answer.

\subsection{Survey results}

\subsubsection{Question 1}

\begin{figure}[h]
  \centering
  \includegraphics[width=0.45\textwidth]{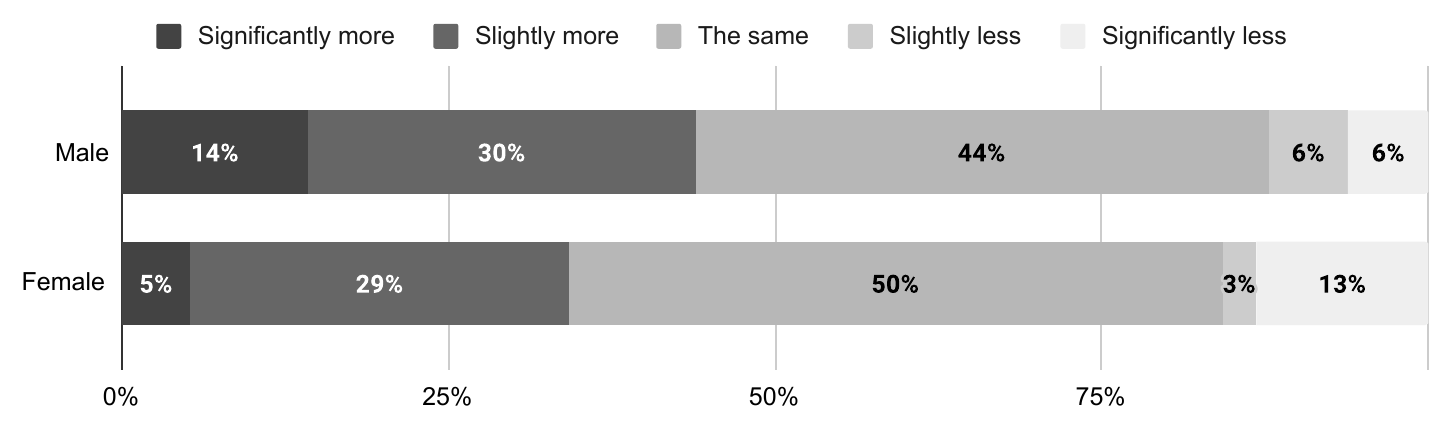}
  \caption{Perceived online forum participation by gender}
  \label{fig:survey_q1}
\end{figure}

We see in Figure \ref{fig:survey_q1} that for Question 1, a higher proportion of males reported having the impression of participating more on the forum in online classes, compared to the female respondents.
A higher proportion of females reported having the impression of participating less on the forum in online classes, or participating the same as in in-person.

Conversely, our analysis of class data suggests that female students are participating more on the forum now that classes are online, or perhaps that males are participating less.
But it seems that the students' perception, at least that of the respondents, is the opposite.
Of course, it is possible that we have self-selection bias in the population of students who chose to complete our survey.
But these responses suggest that female students may have a more modest view of their participation on the forum.

In the free responses to this question, we observe two major trends. The first trend was widely reported by both male and female respondents: now that classes are online, many students are using the forum more because they see it as the only means of communication.
As a male student put it, \textit{"since everything is now online, this provides more incentive to engage with professors and TAs through these platforms such as Discord, Piazza, and CampusWire."}
Another male student mentioned that \textit{"[the] inability to talk to classmates in person has caused [them] to participate slightly more on the forum."}
A female student commented that \textit{"since everything is virtual, the only way to communicate with others is through the internet so \dots class forums are very useful and an essential way to feel slightly more connected with other students when classes are online."}

On the other hand, the second trend was widespread only among female respondents: in online classes, some female students use the forum more because of increased comfort and confidence that comes from their anonymity in the class.
One male student did respond saying that \textit{"in person, [they are] too shy to talk to people, [and that] being online helps a bit."}
However, this was only one of the 24 male students who left free responses explaining why their forum participation had increased, and therefore not enough to establish a trend.
In comparison, of the eight females who left free responses explaining why their forum participation had increased, half of them referenced this reasoning.
One female student replied, \textit{"[they] feel more comfortable while people cannot see [them]"} and another mentioned, \textit{"online classes allow [them] to be anonymous while participating."}

There is a bit of ambiguity here, as anonymous online forum participation is not unique to online classes, and was already an option when the classes we are considering were in-person.
These responses do seem to suggest that some female students have felt increased anonymity now that there is no in-person component to the class, like in-person lecture, and that this overall increase in anonymity is leading to their increase in forum participation.

\subsubsection{Question 2}

\begin{figure}[h]
  \centering
  \includegraphics[width=0.45\textwidth]{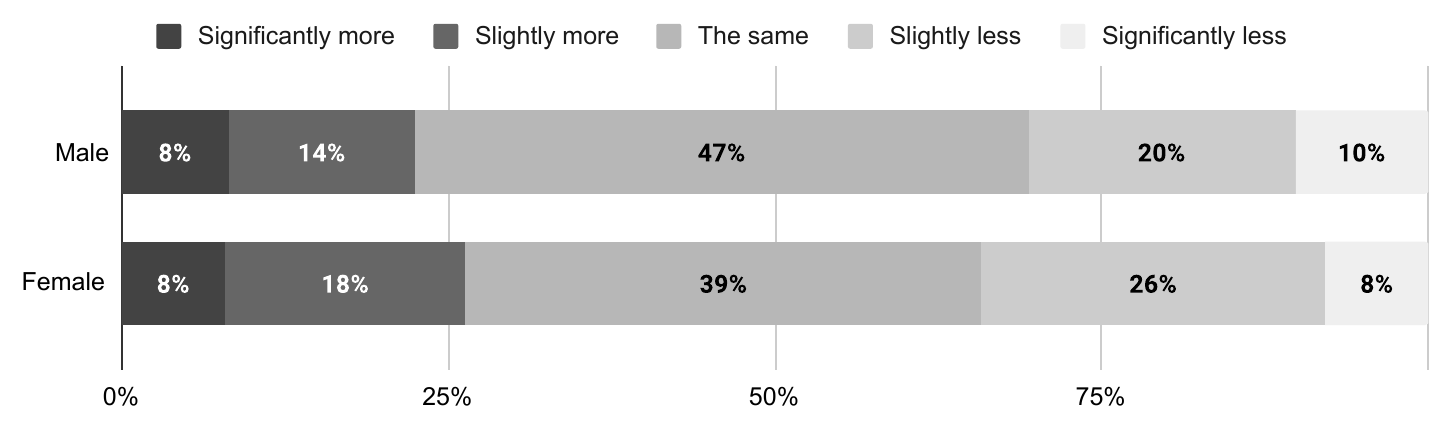}
  \caption{Self-confidence by gender}
  \label{fig:survey_q2}
\end{figure}

We see in Figure \ref{fig:survey_q2} that for Question 2, both the male and female responses were rather normally distributed.
However, a smaller proportion of female respondents reported feeling that their confidence levels had been unaffected by the transition.
Instead, a higher proportion of females reported the shift to have either positively or negatively affected their confidence levels, in comparison to the male respondents.

In the free responses to this question, we observe three major trends. The first trend, shared across all respondents, is that with online classes, many students are less confident in their ability to succeed due to a lack of support and motivation.
One male student commented that \textit{"[they] have less opportunities to ask other classmates or professors for help and advice[, and that they] also feel significantly less engaged when a class is online rather than in person."}
Another male student said, \textit{"[they] don't have a network of people to rely on in the same way [as in in-person classes]."}
A female student reported \textit{"not learning as effectively as [they] would in an in person class because it is much more difficult to stay motivated online."}
Another female commented that \textit{"since there is no in person interaction, it makes [them] feel as if [they're] alone and don't have the support needed."}

On the other hand, the second trend across both male and female students was that in online classes, some students are more confident in their ability to succeed due to the increased accessibility of class material.
A male student made several points, saying that \textit{"[they] appreciate how [they] can take classes and watch lectures at [their] own pace.
When things are in person, there is only one opportunity to listen to a lecture which limits the possibility for a student to ask clarifying questions.
The online content is typically more condensed and easier to access."}
A female student said that for them, \textit{"it's more convenient to watch lectures in [their] own timeline and [having] the ability to pause and take better notes [has] really boosted [their] level of understanding and therefore [their] level of confidence in the class."}

The third trend, which was unique to female respondents, was that some females feel an increased sense of confidence in online classes because there is less opportunity for them to compare themselves to their classmates.
No male respondents echoed this sentiment.
One female student mentioned that they \textit{"feel more comfortable when [they] don't see all the students, who are younger, smarter and better than [themselves].
[In online classes, they aren't] scared [of being] judged anymore."}
Another female mentioned that \textit{"something that used to lower [their] confidence in [their] cs abilities [was], during in person classes, when someone would raise their hand and ask questions that [they] had absolutely no idea about."}

In this last trend, we notice that female respondents already bring up comparison to peers without being prompted.
However, it is not until our next question, Question 3, that we explicitly asked students to consider the amount that they compared themselves to their classmates.
This supports our initial intuition that students' attitudes towards their peers may have been a significant driver of the behavioral changes we observed on the forum.

\subsubsection{Question 3}

\begin{figure}[h]
  \centering
  \includegraphics[width=0.45\textwidth]{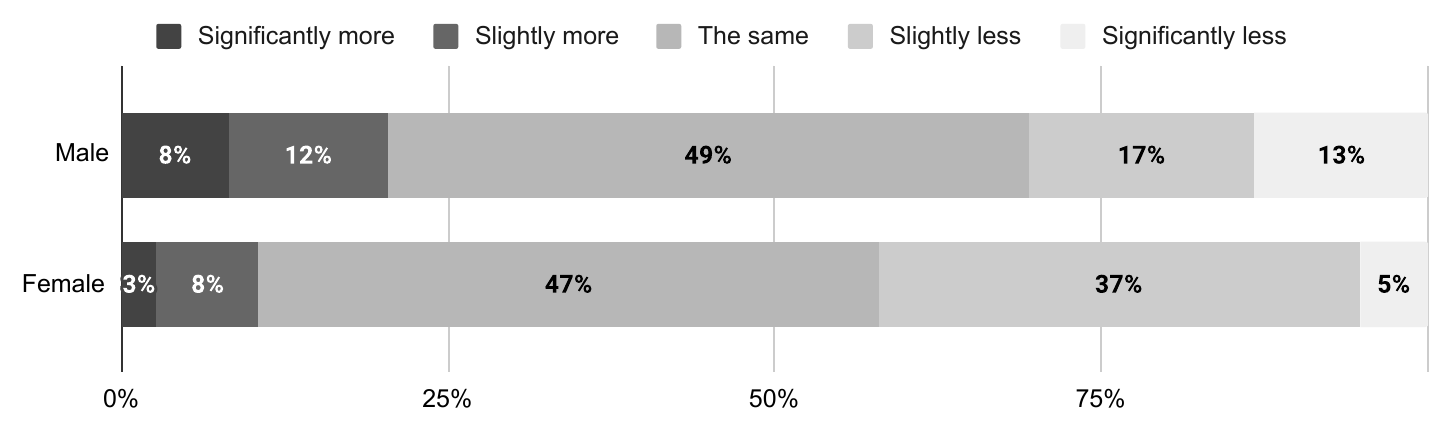}
  \caption{Attitude towards peers by gender}
  \label{fig:survey_q3}
\end{figure}

We see in Figure \ref{fig:survey_q3} that for Question 3, a higher proportion of female respondents reported comparing themselves to their peers less now that classes were online.

In the free responses to this question, we observe three major trends.
With classes online, many male and female students compare themselves to their peers less because not seeing them has created a feeling of indifference.
Or, as one male student put it, \textit{"[they're] all just Zoom names to me now :)"}.
Another male student explained, \textit{"it's harder for [them] to see how prepared [or] confident [their classmates] are, and [that they generally] think less about them because [they] don't see them personally each and every single day now."}
A female student commented that \textit{"with classes online, [they're] not surrounded by [their] classmates everyday so [they're] less likely to take note of the differences [or] compare [themselves] with them."}

The second trend is the first trend unique to male students: in online classes, some males report comparing themselves to their classmates more, because there is more room to speculate and make assumptions about their progress.
This reasoning was not echoed among female respondents.
One male explained that \textit{"since [they] feel less engaged in class, [they] worry that [they are] not learning as much information as other students and thus feel less prepared for quizzes, exams, and assignments when compared to other[s]."}
Another commented that \textit{"[they] don't get to see [their classmates] in person so [they] construct a version of them in [their] head that may or may not be true and is only based on what [they] see online."}
Another male student mentioned they \textit{"feel that at home, [they] are able to slack off more while [their] peers are able to get ahead in assignments, projects, and extracurriculars."}

Conversely, the final trend was unique to female respondents: in online classes, some females compare themselves to their peers less because they feel more comfortable not interacting with them in an in-person lecture environment. This was not echoed by any male respondents.
One female commented that \textit{"being in in-person lectures can make the class seem more intimidating [and] intense.
You don't experience that environment when lectures are online."}
Another female explained that \textit{"with in-person classes, there is more opportunity for participation.
Those who ask questions are usually more outspoken and more knowledgeable than [they are], so [they] feel more self-conscious of [their] own abilities and knowledge.
These questions [in lecture] are sometimes very specific and maybe even unrelated to the class.
With the forum, the questions are usually more focused and [they] do not feel as self-conscious when answering them."}

This is an interesting difference between male and female students.
It seems like for some female students, seeing and interacting with classmates in person makes them question their own abilities, whereas for some male students, this interaction actually makes them more confident.
Therefore, the in-person lecture component of CS classes may be more favorable to some male students, and less favorable to some female students, due to the class format's effect on students' self-perception.
This is telling of what kind of student benefits from the classroom climate of CS classes.
As classroom climate can impact a student's self-perception, this could explain why in in-person classes, we even see reduced female participation in spaces outside of the in-person components, namely, on the forum.

\subsubsection{Non-binary responses}

While the small sample size of non-binary students prevents us from drawing statistically significant conclusions, it is still interesting to consider this group's responses.
The two non-binary students responded neutrally to all Likert scale questions; they reported that the shift to online classes had no impact on their forum participation, confidence in their CS abilities, or comparison to their classmates.
Neither of the students left free responses.

\section{Limitations}

We obtained official gender identification information for each student's course grade from the university registrar, which uses a binary gender classification.
To avoid an oversimplified gender spectrum in future studies, students could self-report their gender identity.

Naturally, the duration of the pandemic limited the size of our online class dataset.
Analyzing data from online classes taught before the pandemic would allow us to gauge the consistency of the observed trends.

The same male instructor taught all analyzed classes.
Future research could investigate whether we see consistent trends in similar classes taught under different instructors.

Alongside the shift from in-person to online, the analyzed classes also moved from a synchronous to a primarily asynchronous format.
This secondary shift in format may also impact how students of different gender identities participate in class.

\section{Conclusion}

As almost all aspects of CS3 and CSOS remained consistent between the in-person and online offerings (\eg the same professor, the same class structure), this allows us to isolate and study the potential impacts of the only few but significant changes: the transition from an in-person to online delivery of lectures, discussions, and office hours, and the transition from synchronous to asynchronous lectures.
Our data analysis finds a correlation between the shift from an in-person to a primarily asynchronous online class format and the closure of the gender gap in forum participation class-wide, which, with the removal of lecture participation, is now the most public form of participating.
These observations differ from existing research findings on participation in in-person classes, which found that the most public forms of participation are largely male-dominated \cite{1,3,9}.

Student survey responses suggest that female students are less likely to compare themselves to their peers in online classes, potentially contributing to increased female participation on the forum.
Conversely, our collected responses suggest that male students are more likely than females to compare themselves to their peers when they do not see them in person.
This observation suggests that the classroom climate in in-person settings could benefit male students.

The solution to alleviating the class participation gender gap in CS classes is not to remain online forever.
However, these observed differences in participation between in-person and online classes can help us better pinpoint what we need to address to mitigate the gap when we return to an in-person format.
This data suggests that observed female participation habits in typical in-person classes are not inherent gender differences but rather a product of the environment.
Future work should explore the causes of these behavioral differences (\ie why some females report feeling more confident online than in person, why some males report the opposite) and experiment with ways to ensure in-person computer science classes promote equity for students of all gender identities.

\balance
\bibliographystyle{ACM-Reference-Format}
\bibliography{participation}

\end{document}